Full Paper

# Capillary nanostamping with spongy mesoporous silica stamps

*Mercedes Schmidt, Michael Philippi, Maximilian Münzner, Johannes M. Stangl, René Wieczorek, Wolfgang Harneit, Klaus Müller-Buschbaum, Dirk Enke, Martin Steinhart\**

M. Schmidt, M. Philippi, Prof. M. Steinhart
Institute of Chemistry of New Materials, Universität Osnabrück, 49069 Osnabrück, Germany
martin.steinhart@uos.de

M. Münzner, Prof. D. Enke
Institute of Technical Chemistry, Universität Leipzig, 04103 Leipzig, Germany

J. M. Stangl, Prof. K. Müller-Buschbaum
Institute of Inorganic Chemistry, Universität Würzburg, 97074 Würzburg, Germany

R. Wieczorek, Prof. W. Harneit
Department of Physics, Universität Osnabrück, 49069 Osnabrück, Germany

**Abstract:** Classical microcontact printing involves transfer of molecules adsorbed on the outer surfaces of solid stamps to substrates to be patterned. We prepared spongy mesoporous silica stamps that can be soaked with ink and that were topographically patterned with arrays of submicron contact elements. Multiple successive stamping steps can be carried out under ambient conditions without ink refilling. Lattices of fullerene nanoparticles with diameters in the 100 nm range were obtained by stamping $C_{60}$/toluene solutions on perfluorinated glass slides partially wetted by toluene. Stamping an ethanolic 1-dodecanethiol solution onto gold-coated glass slides yielded arrays of submicron dots of adsorbed 1-dodecantethiol molecules, even though macroscopic ethanol drops spread on gold. This outcome may be related to the pressure drop across the concave ink menisci at the mesopore openings on the stamp surface counteracting the van der Waals forces between ink and gold surface and/or to reduced wettability of the 1-dodecanethiol dots themselves by ethanol. The chemical surface heterogeneity of gold-coated glass slides functionalized with submicron 1-dodecanethiol dots was evidenced by dewetting of molten polystyrene films eventually yielding ordered arrays of polystyrene nanoparticles.

# 1. Introduction

Chemical and topographical patterning of substrates by contact lithography has been the focus of significant research efforts for more than two decades. Contact-lithographic methods can be separated into scanning probe-based and stamp-based approaches. Scanning-probe lithography, such as dip pen lithography[1–3] and methods involving liquid supply via cantilever tips,[4–7] as well as related micropipetting techniques,[8] are commonly serial low-throughput methods (processing time for 100 µm x 100 µm > several minutes). By contrast, stamp-based nanolithography enables high-throughput modification of large substrate areas. Examples include soft lithography with elastomeric stamps,[9,10] microcontact printing,[11,12] polymer pen lithography,[2,3,13] electrochemical lithography,[14] as well as different types of capillary force lithography[15,16] and wet lithography.[17] Conventional nanolithography configurations based on the use of solid stamps only allow the transfer of molecules adsorbed on the stamp surface to the substrate to be patterned and often require controlled atmospheric conditions, frequent readsorption of ink or long cycle times. Here, we evaluate the potential of capillary nanostamping with spongy mesoporous silica stamps to mitigate the drawbacks of the conventional stamping methods mentioned above. Mesoporous silica stamps may enable *in situ* ink supply from an external ink reservoir during stamping. Hence, continuous stamp operation is possible without the need of interruptions for the readsorption of ink, as it is required in the case of conventional solid stamps. Alternatively, the mesopore systems of the mesoporous silica stamps themselves may be used as internal ink reservoirs. Thus, multiple stamping cycles can be performed without deterioration of the quality of the stamped pattern. In contrast to conventional stamping methods relying on solid stamps that often require controlled atmospheric conditions, capillary nanostamping can conveniently be carried out under ambient conditions, and larger amounts of functional ink components can be transferred to counterpart substrates. We investigated two limiting cases of capillary nanostamping using two model inks. Stamping solutions of fullerene $C_{60}$ in toluene onto perfluorinated glass slides corresponds to a scenario in which the non-volatile ink component $C_{60}$ has a higher affinity to itself than to the substrate to be stamped. Therefore, regular arrays of $C_{60}$ nanoparticles are obtained. $C_{60}$ is a model for endohedral fullerenes.[18] Moreover, $C_{60}$ is being considered as one of the best electron acceptors for organic solar cells.[19] Morphology design via controlled assembly of $C_{60}$ may pave the way for efficient phase-separated donor/acceptor systems.[20] However, while $C_{60}$ nanoparticles have gained significant interest[21,22] controlled assembly of fullerenes[23] has remained challenging. Stamping ethanolic 1-dodecanethiol solutions onto gold surfaces corresponds to a scenario in which the non-volatile ink component 1-dodecanethiol exhibits the strongest affinity to the substrate to be stamped. We show that, even though the solvent ethanol spreads on gold, capillary nanostamping of ethanolic 1-dodecanethiol solutions yields ordered arrays of 1-dodencanethiol submicron dots. This outcome is not only of interest because 1-dodecanethiol is representative of the broad variety of functional thiols[24] used for the generation of chemically heterogeneous surfaces. We also demonstrate that arrays of submicron 1-dodecanethiol dots generated by capillary nanostamping on gold template the dewetting of molten polystyrene (PS).

# 2. Results and discussion
## 2.1 Design of mesoporous silica stamps

For capillary nanostamping we used monolithic mesoporous silica stamps entirely penetrated by continuous, spongy mesopore systems because of their chemical stability against organic inks. We synthesized the mesoporous silica stamps by sol-gel chemistry; to generate silica gel bodies topographically patterned with arrays of contact elements, the sols were molded against macroporous silicon (mSi) (Supporting Figure S1).[25,26] The mSi contained hexagonal arrays of macropores with a center-to-center distance of 1.5 µm and a macropore depth of 600 nm. The diameter of the macropores amounted to 900 nm at their openings and to 600 nm at their bottoms. We could non-destructively separate the gelated mesoporous silica stamps from the mSi because the mSi surface, initially consisting of native silicon oxide, was modified with alkylated silane. To test whether synthetic routes based on the common tetrafunctional silica precursor tetraethyl orthosilicate (TEOS) yield applicable mesoporous silica stamps, we adapted a synthetic protocol reported by Babin et al.[27] The mesoporous silica stamps obtained in this way (Supporting Figure S2) with mesopore diameters of 10-20 nm (Supporting Figure S3) were too brittle and did not withstand ink infiltration. By adapting a modified synthetic approach[28] we obtained mesoporous silica stamps consisting of TEOS-derived silica networks that were reinforced by incorporation of titanium atoms. While their mechanical stability was sufficient, they suffered from broad mesopore diameter distributions (the mesopore diameters ranged from 20-60 nm; Supporting Figure S3) and from too wavy surfaces (Supporting Figure S4).

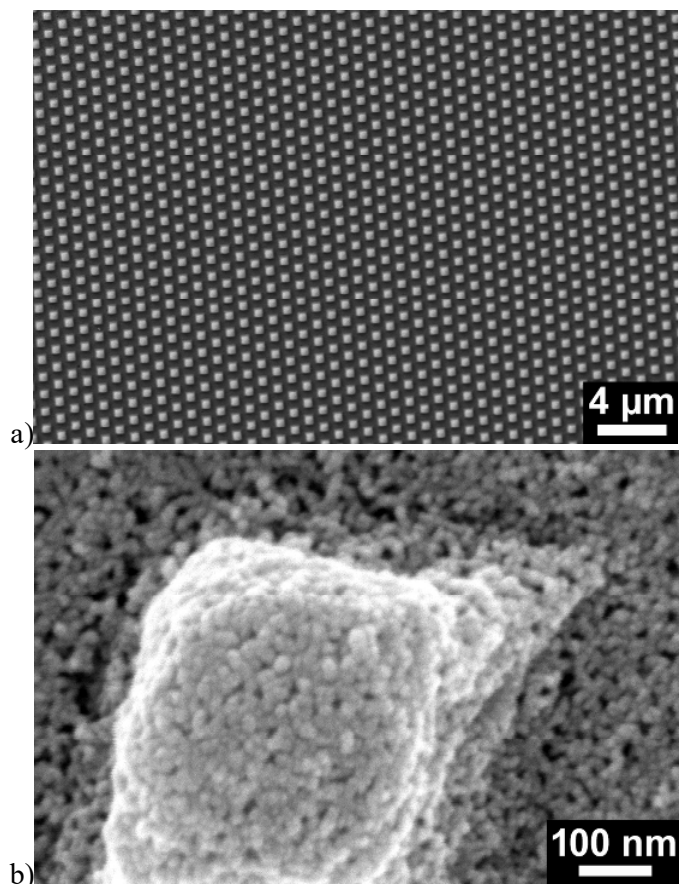

**Figure 1.** Scanning electron microscopy images of mesoporous silica stamps prepared by molding MTMS-containing sols previously introduced by Kanamori et al.[29] against silanized mSi. a) Top view; b) detail.

Eventually, we obtained hydrophobic silica networks containing incorporated methyl groups that were usable as mesoporous silica stamps (Figure 1a) by molding sols containing the trifunctional silica precursor methyltrimethoxysilane (MTMS)[29] against silanized mSi. As outlined by Kanamori et al., silica aerogels consisting of rigid tetrafunctional siloxane networks, as obtained when TEOS is used as silica precursor, are brittle and prone to breakage. MTMS, however, forms trifunctional siloxane networks that exhibit lower crosslinking densities and, consequently, higher flexibility than tetrafunctional siloxane networks.[29] The higher flexibility of MTMS-derived trifunctional siloxane networks reduces the extent of undesired deformations of the mesoporous silica stamps during gelation. The lateral shrinkage in the course of gelation amounted to 25-35 %, while the height shrunk by 15-25 %. The surfaces of the mesoporous silica stamps were topographically patterned with regular hexagonal arrays of conical contact elements. The contact elements had a base diameter of ~600 nm, a tip diameter of ~400 nm and a height of ~500 nm (Figure 1b). The nearest neighbor distance between adjacent contact elements amounted to ~1 μm. Nitrogen sorption measurements (Supporting Figure S5a and Table S1) revealed a BET surface area of 530 m$^2$/g. The shape of the nitrogen sorption isotherm could be assigned to type IV with a H1 hysteresis loop. Generally, H1 hysteresis loops are characteristic of materials with relatively uniform cylindrical mesopores.[30] Here, the steep and narrow H1 hysteresis loop indicates that the necks and the nodes of the spongy-continuous mesopore system are uniform in size – the average pore diameter derived from the adsorption branch of the nitrogen sorption isotherm calculated by the BJH method amounted to 31 nm (Supporting Figure S5b). By contrast, a mesopore system consisting of narrow necks and wider nodes would yield a H2 hysteresis loop. The results of the nitrogen sorption measurements obtained here are in line with results reported by Yun et al.,[31] who synthesized silica microspheres using MTMS as silica precursor. The total pore volume of 2.2 mL/g suggests that an empty mesoporous silica stamp with a mass of ~90 mg may absorb ~200 μL ink.

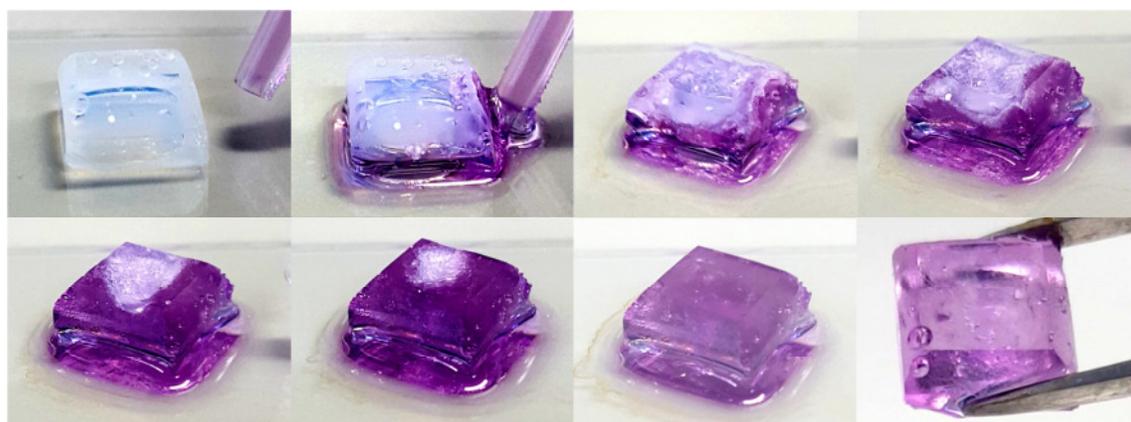

**Figure 2.** Photographs showing the imbibition of a mesoporous silica stamp with a solution of $C_{60}$ in toluene (pink). The mesoporous silica stamp with edge lengths of ~5 mm is located on a glass slide in such a way that the contact surface of the mesoporous silica stamp topographically patterned with the contact elements points upwards. Prior to imbibition, the empty mesoporous silica stamp is opaque and colourless. The $C_{60}$/toluene solution dropped onto the glass slide close to the mesoporous silica stamp invades the latter. The portions of the mesoporous silica stamp filled with $C_{60}$/toluene solution are transparent-pink.

## 2.2 Capillary nanostamping with mesoporous silica stamps

The mesopore system enables either continuous or discontinuous ink supply from the backside of the mesoporous silica stamps to the opposite contact surface patterned with the contact elements. For example, we imbibed a mesoporous silica stamp with a solution of 1 mg $C_{60}$ per 1 mL toluene (Figure 2; Supporting Movie 1). The mesoporous silica stamp was located on a glass slide in such a way that the contact surface topographically patterned with the contact elements pointed upwards. The pink $C_{60}$/toluene solution was dropped onto the glass slide close to the position of the mesoporous silica stamp. The series of photographs seen in Figure 2 and Supporting Movie 1 evidence that capillary action sucked the $C_{60}$/toluene solution into the mesopore system of the mesoporous silica stamp. While the empty portions of the mesoporous silica stamps appear opaque-colorless, the portions of the mesoporous silica stamp filled with $C_{60}$/toluene solution appear transparent-pink. The imbibition front moved upwards until the entire mesopore system was filled with $C_{60}$/toluene solution. Also, an ethanolic solution of 1-dodecanethiol readily imbibed the mesoporous silica stamps (Supporting Figure S6 and Supporting Movie 2).

The MTMS-derived mesoporous silica stamps withstood the mechanical impact during stamping. A potential problem to be overcome is misalignment between the mesoporous silica stamps and the substrate to be patterned. Therefore, we glued the mesoporous silica stamps onto elastomeric PDMS films with a thickness of ~3 mm that were in turn connected to a stamp holder made of steel. When the ink-filled mesoporous silica stamps (Figure 3a) are brought into contact with the substrates to be stamped, the elastic deformability of the PDMS layer compensates misalignment between stamp holder, mesoporous silica stamp and substrate. As discussed below, the contact time between stamp and substrate is a parameter that needs to be adjusted to the specific capillary nanostamping process.

The ease with which the mesoporous silica stamps can be filled with ink from external ink reservoirs suggests that ink can easily be supplied during stamp operation. Thus, the mesoporous silica stamps can continuously be kept wet during use; no interruption of stamp operation for refilling with ink is necessary. If the mesoporous silica stamps are not connected to external ink reservoirs, their mesopore systems may serve as internal ink reservoirs. Under ambient conditions ink-filled mesoporous silica stamps can conveniently be handled and multiple capillary nanostamping cycles can be carried out without deterioration of the quality of the stamped patterns. For example, as discussed below more than ten capillary nanostamping cycles can be performed with ethanolic ink stored in the mesopore system of the mesoporous silica stamps without controlled atmospheric conditions. The limiting factor for the number of stamping cycles that can be performed under ambient conditions without connecting the stamps to external ink reservoirs is the evaporation of volatile ink components such as solvents. Despite their enhanced mechanical robustness, even the MTMS-derived mesoporous silica stamps do not withstand complete drying under ambient conditions. However, when not in use the mesoporous silica stamps can be kept wet to maintain their functionality by storing them in immersed in a liquid. It is also possible to wash residual ink out of the mesoporous silica stamps with suitable solvents (Supporting Figure S7).

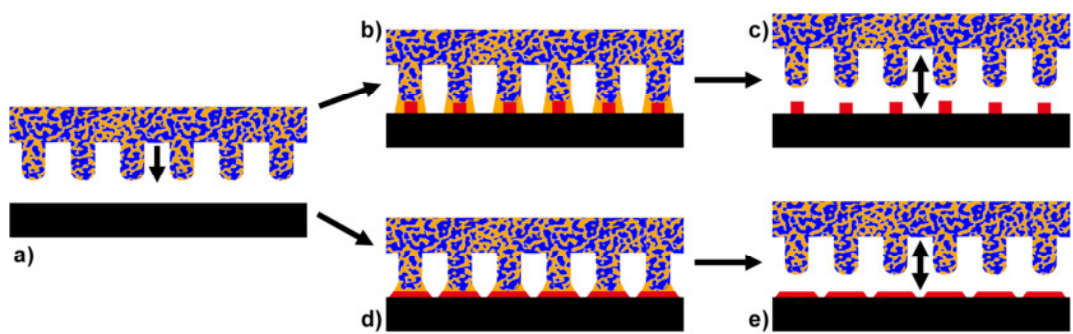

**Figure 3.** Capillary nanostamping. a) A mesoporous silica stamp (blue) soaked with ink (orange) is approached to a substrate to be patterned (black). b),d) Once the gap between substrate and mesoporous silica stamp is small enough, liquid bridges form between the contact elements of the mesoporous silica stamp and the substrate. b) If $C_{60}$/toluene solution is stamped on perfluorinated glass slides, the ink is only partially wetting. Solvent evaporation draws additional ink into the liquid bridges so that $C_{60}$ enriches and crystallizes into nanoparticles (red). c) After detachment of the mesoporous silica stamps, the $C_{60}$ nanoparticles remain at the former position of the contact elements. d) If 1-dodecanethiol/ethanol solution is stamped onto gold-coated glass slides, submicron 1-dodecanethiol dots (red) form at the position of the contact elements by adsorption of 1-dodecanethiol onto the gold surface. e) After detachment of the mesoporous silica stamp a chemically patterned surface is obtained.

## 2.3 Capillary nanostamping of $C_{60}$ nanoparticles

We stamped solutions of 1 mg $C_{60}$ per 1 mL toluene onto glass slides perfluorinated with 1H,1H,2H,2H-perfluorodecyltrichlorosilane (FDTS). We brought the mesoporous silica stamps filled with $C_{60}$/toluene solution manually into contact with FDTS-functionalized glass slides for ~5 s. The contact angle of toluene on FDTS-functionalized glass slides amounted to

~71°. Thus, stable liquid ink bridges formed between the contact elements of the mesoporous silica stamps and the FDTS-modified glass slide (Figure 3b), whereas ink spreading was thermodynamically prevented. A contact time of 5 s between mesoporous silica stamp and FDTS-modified glass slide was necessary to enable transport of additional ink into the liquid bridges driven by evaporation of the volatile ink component toluene. In this way the nonvolatile ink component $C_{60}$ enriched within the liquid bridges and finally crystallized into nanoparticles. Thus, ordered arrays of $C_{60}$ nanoparticles remained on the FDTS-coated glass slides after the retraction of the mesoporous silica stamps (Figure 3c). Figure 4a shows a scanning electron microscopy (SEM) image of a $C_{60}$ nanoparticle array taken with a secondary electron chamber detector emphasizing topography. In the SEM images seen in Figure 4b) and c) taken with an in-lens detector the areas of the $C_{60}$ nanocrystals having low work functions appear bright while the surrounding FDTS-modified surface of the glass slide (high work function) appears dark. Large-area SEM images of arrays of $C_{60}$ nanoparticles are shown in Supporting Figure S8. The arrangement on the FDTS-coated glass slides nearly perfectly replicated the arrangement of the contact elements of the mesoporous silica stamp. The $C_{60}$ nanoparticles had an average diameter of 123 nm ± 22 nm (Supporting Figure S9) and were, therefore, smaller than the contact elements of the mesoporous silica stamps. The height of the $C_{60}$ nanoparticles amounted to 50-70 nm (Supporting Figure S10).

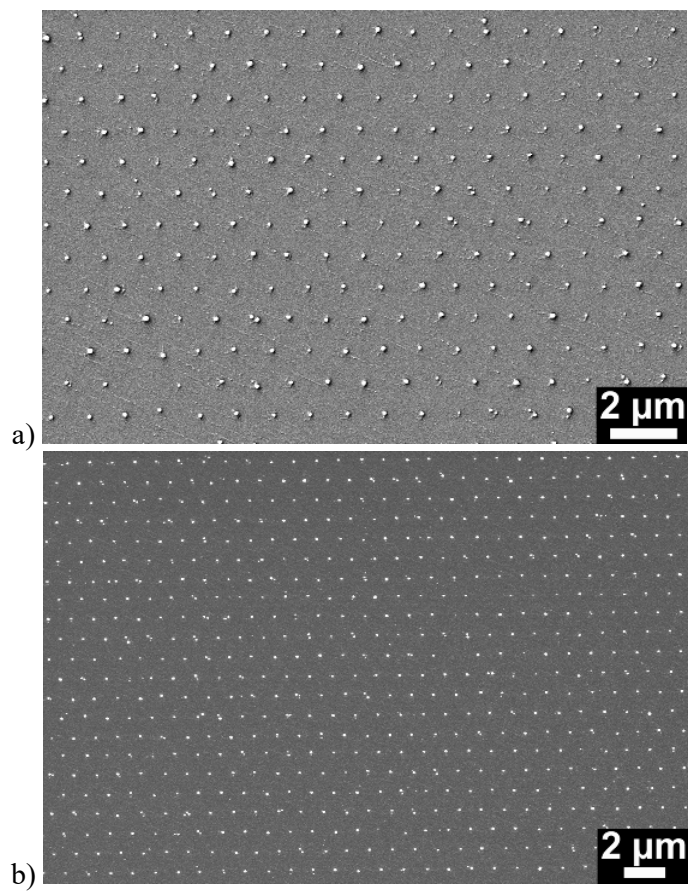

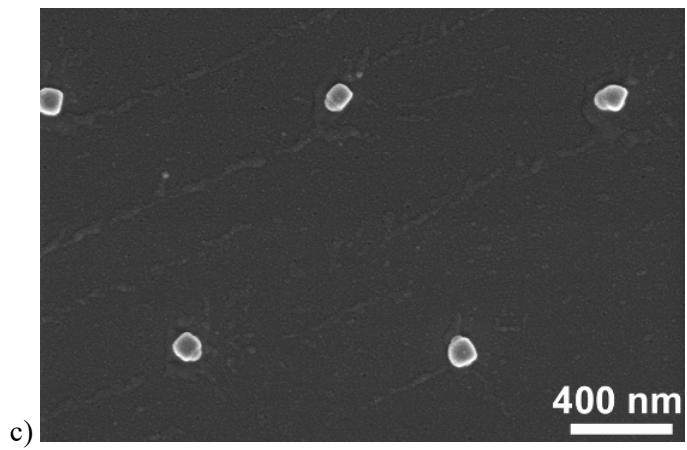
c)

**Figure 4.** SEM images of arrays of $C_{60}$ dots generated by capillary nanostamping on a FDTS-modified glass slide. Overview recorded a) with a secondary electron chamber detector and b) with an in-lens detector; c) detail taken with an in-lens detector.

## 2.4 Capillary nanostamping of submicron 1-dodecanethiol dots

In a second experiment, we stamped ethanolic 1-dodecanethiol solution onto gold-coated glass slides. Formation of capillary bridges during the contact time of ~1 s resulted in adsorption of 1-dodecanethiol onto the gold-coated glass slides at the positions of the contact elements of the mesoporous silica stamps (Figure 3d) owing to the strong affinity of thiol groups to gold surfaces.[32] After detachment of the mesoporous silica stamps, arrays of submicron 1-docecanethiol dots were obtained (Figure 3e). To enhance the SEM contrast between the 1-dodencanethiol dots and the surrounding gold surface, the thickness of the unprotected gold layer separating the stamped 1-dodecanethiol dots was reduced from ~30 nm to ~10 nm by wet-chemical etching (cf. Experimental). Thus, topographic contrast allowed the identification of 1-dodecanethiol dots in SEM images captured with a chamber secondary electron detector (Figure 5a). Like the $C_{60}$ nanoparticles, the 1-docecanethiol dots formed a hexagonal lattice replicating the lattice formed by the contact elements of the mesoporous silica stamps (Supporting Figure S11 shows a large-area image). Figure 5b shows a detail of Figure 5a recorded with an in-lens detector. The areas of the 1-dodecanethiol dots having high work functions appear dark; the surrounding of the 1-dodecanethiol dots consisting of residual gold with low work function appears bright (as compared to the $C_{60}$ nanoparticle arrays on FDTS-modified glass slides the contrast is inverted). Image analysis of a SEM image displaying 1312 1-dodecanethiol dots revealed a nearest-neighbor distance of 1.2 µm, an average dot diameter of 756 nm ± 26 nm (Supporting Figure S12a) and an average dot area of 0.449 µm$^2$ ± 0.031 µm$^2$. Thus, in contrast to the $C_{60}$ nanoparticles the 1-dodecanethiol dots were larger than the contact elements of the mesoporous silica stamps. The circularity $C = 4\pi * area/perimeter^2$ and the aspect ratio (ratio long axis to short axis) of an object are geometric descriptors quantifying the deviation of the object's contour from that of an ideal circle, for which aspect ratio and circularity equal 1. For the evaluated 1-dodecanethiol dots we obtained an average circularity of 0.94 ± 0.03 (Supporting Figure S12b) and an average aspect ratio of 1.09 ± 0.04 (Supporting Figure S12c). The slight deviations from ideal circular shape may be rationalized by slight shear during stamping or by aberration. Capillary nanostamping can be conducted multiple times under ambient conditions without the need to refill the mesoporous silica stamps with ink. For example, we successively stamped 1-dodecanethiol dots onto 10 different gold-coated glass slides without ink refilling. Figure 5c shows an array of 1-dodecanethiol dots resulting from the tenth consecutive capillary nanostamping cycle, revealing no differences between the arrays of 1-dodecanethiol dots obtained in the first and in the tenth successive capillary nanostamping cycle.

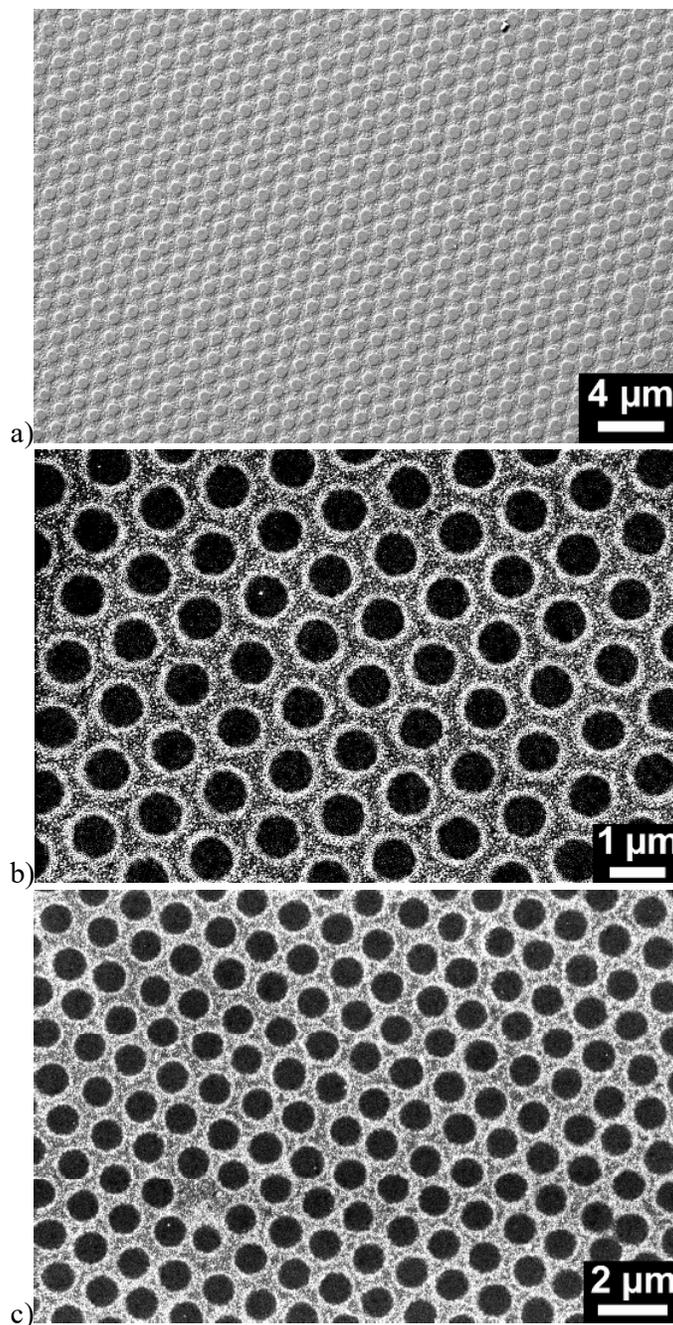

**Figure 5.** SEM images of arrays of 1-dodecanethiol dots generated by capillary nanostamping on gold-coated glass slides after partial etching of the gold separating the 1-dodecanethiol dots. a) Overview after the first stamping cycle acquired with a secondary electron chamber detector. b) Detail of panel a) acquired with an in-lens detector. c) Array of 1-dodecanethiol dots prepared in the 10$^{th}$ successive stamping cycle without refilling the mesoporous silica stamp with ink (image was taken with an in-lens detector).

The mean diameter of the 1-dodecanethiol dots is by a factor of 1.75 larger than the diameter of the contact elements of the mesoporous silica stamps. It is remarkable that discrete submicron 1-dodecanethiol dots were obtained despite the fact that macroscopic ethanol drops spread on the gold-coated glass slides (corresponding to a contact angle of 0°). This outcome may be rationalized as follows. The ink is drawn out of the contact elements of the mesoporous silica stamps by van der Waals interactions between ink and gold-coated glass slide. However, the entire surface of the mesoporous silica stamps exhibits mesopore openings where the ink forms concave menisci. Because of the concave curvature of the menisci the ink pressure $p_{ink}$ is lower than the air pressure outside the mesoporous silica stamp $p_{outside}$ (Figure 6). The pressure drop across the menisci $\Delta p = p_{outside} - p_{ink}$ referred to as Laplace pressure[33] counteracts the attractive van der Waals forces between ink and gold-coated glass slide. Therefore, it is reasonable to assume that the presence of open spongy mesopores in the silica stamps impedes spreading of ethanol on the gold-coated glass substrates. On the other hand, the contact angle of ethanol on gold-coated glass slides immersed for 5 s into a solution of 2 mM 1-dodecanethiol in ethanol increased to ~34°. Hence, the adsorption of 1-dodecanethiol on the gold-coated glass substrates itself in the course of capillary nanostamping may suppresses ink spreading.

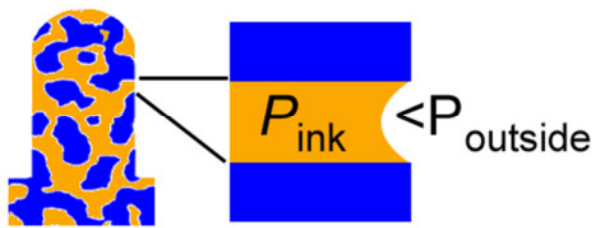

**Figure 6.** Schematic diagram showing a concave meniscus formed by ink (orange) at a mesopore opening of a mesoporous silica stamp. The pressure of the ink $p_{ink}$ is smaller than the air pressure $p_{outside}$. The pressure difference across the meniscus $\Delta p = p_{outside} - p_{ink}$ is the Laplace pressure.

## 2.5. Dewetting of polystyrene on gold films modified with stamped arrays of submicron 1-dodecanethiol dots

Chemically heterogeneous surfaces have attracted significant interest.[34] Surface patterns of areas with chemically different properties template the dewetting of polymeric liquids deposited on the chemically heterogeneous surfaces.[35,36] Dewetting is, therfore, an ideal means to probe the chemical heterogeneity of the gold-coated glass slides functionalized with arrays of submicron 1-dodecanethiol dots by capillary nanostamping. As model liquid for the dewetting experiments we selected molten PS because dewetting of PS has been intensively studied.[37,38] We spin-coated solutions of PS in chloroform onto Au-coated glass slides patterned with 1-dodencanethiol dots. Subsequently, the samples were heated for 30 minutes to 170 °C, a temperature where PS is liquid. PS selectively wets gold because the polarizability of gold is higher than that of 1-dodecanethiol. Thus, the attractive London dispersion interactions between PS and gold are stronger than between PS and 1-dodecanethiol. The dewetting velocities slightly varied in different sample regions so that we could acquire snapshots of different wetting stages. In some parts of Figure 7a a continuous PS film still exists. Also seen, however, are circular holes in the PS film at the positions of 1-dodecanethiol dots. Many of these holes have smaller areas than the 1-dodecanethiol dots, suggesting that dewetting is initiated by nucleation of holes in the PS film at positions where the PS film covers 1-dodecanethiol dots. Dewetting then proceeds by growth of the holes and formation of thickened PS rims surrounding the growing holes. The thickened rims move outwards until they impinge on each other in areas where gold is exposed to the PS. As a result, the PS leaves the 1-dodecanethiol dots blank and forms a honeycomb pattern that coincides with the exposed gold surface separating the discrete 1-dodenanethiol dots (Figure 7b). As indicated in Figure 7c, the honeycomb pattern further decomposes into discrete triangular PS nanoparticles located in the gaps between three adjacent 1-dodecanethiol dots. Eventually, ordered arrays of polystyrene nanoparticles are obtained (Figure 7d). This is obvious from the Fourier spectrum of an in-lens SEM image showing the same image area as Figure 7d (Supporting Figure S13).

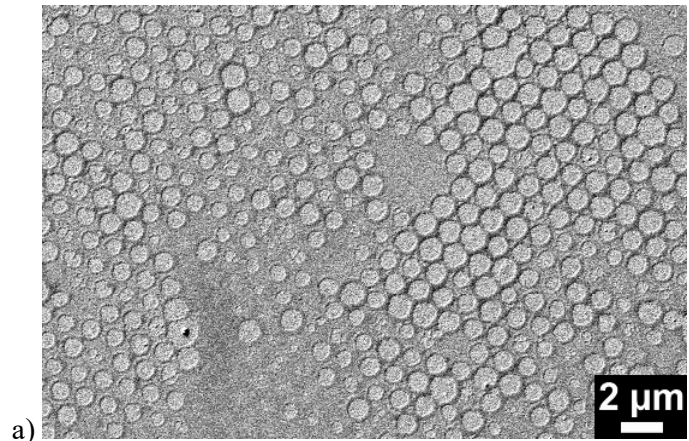

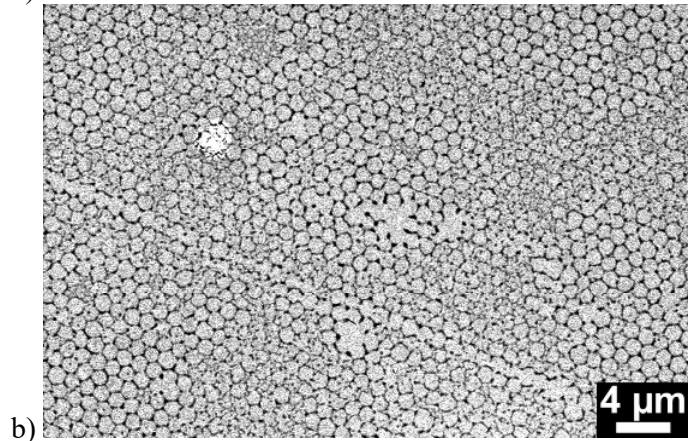

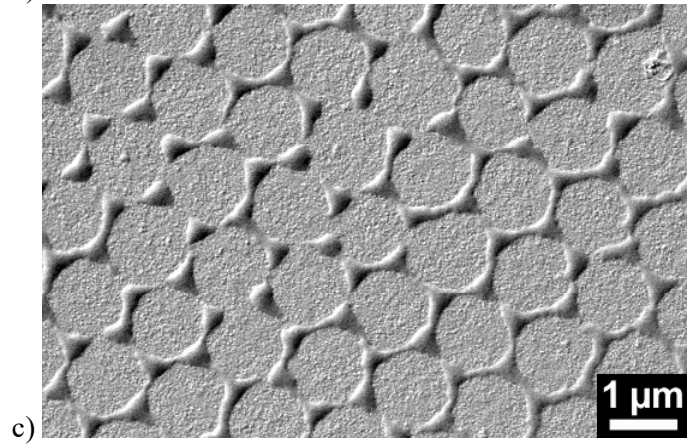

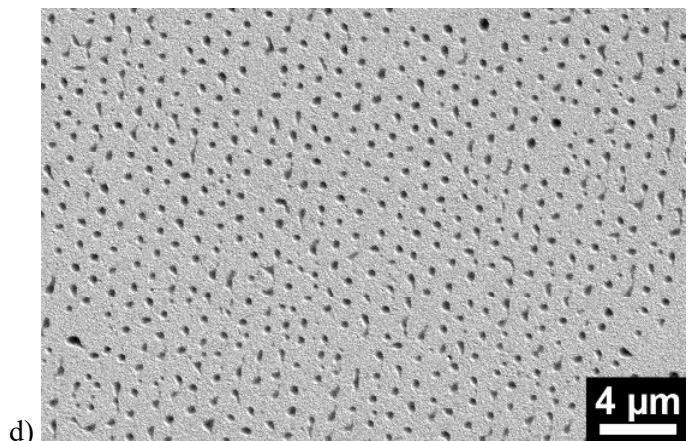
d)

**Figure 7.** SEM images of dewetted PS films on Au substrates patterned with submicron 1-dodecanethiol dots. The PS films had a), c) an initial thickness of ~25 nm and were obtained by spin-coating 0.5 wt-% PS in chloroform or b), d) an initial thickness of ~20 nm and were obtained by spin-coating 0.05 wt-% PS in chloroform. Note that the post-stamping etching step to partially remove the gold for SEM contrast enhancement was *not* performed here. a) Holes at the positions of 1-dodecanethiol dots. b) Honeycomb pattern. c) Detail of a honeycomb pattern. d) Array of discrete PS nanoparticles. All images were acquired with a secondary electron chamber detector.

## 3. Conclusions

Classical microcontact printing is often carried out under controlled atmospheric conditions and involves the use of solid stamps, onto the outer surface of which the species to be stamped needs to be adsorbed. We have investigated capillary nanostamping with silica stamps penetrated by spongy continuous mesopore systems as an alternative approach to surface patterning under ambient conditions. During stamp operation ink can continuously be supplied through the mesopore systems to the contact surfaces of the mesoporous silica stamps. Alternatively, the mesopore systems of the mesoporous silica stamps themselves can be employed as internal ink reservoirs. Even without ink supply from external ink reservoirs capillary nanostamping can be repeated multiple times without the need to refill the mesoporous silica stamps with ink and without deterioration of the quality of the stamped pattern. The mesoporous silica stamps were topographically patterned with arrays of contact elements ~400 nm in diameter. Upon approach to a counterpart substrate, attractive van der Waals interactions between ink and substrate result in the formation of liquid bridges at the positions of the contact elements. In a first example, $C_{60}$/toluene inks were stamped onto perfluorinated glass slides that are partially wetted by the ink. Evaporation of toluene draws fresh ink into the liquid bridges in which the $C_{60}$ enriches and crystallizes. The mean diameter of the $C_{60}$ nanoparticles formed at the positions of the contact elements amounting to 123 nm ± 22 nm was smaller than the contact area of the contact elements. In a second example, stamping of 1-dodecanethiol/ethanol solutions onto gold-coated glass slides yielded arrays of submicron dots consisting of adsorbed 1-dodecanethiol molecules. The 1-dodecanethiol dots had a mean diameter of 756 nm and were, therefore, larger than the contact elements. Remarkably, discrete 1-dodecanethiol dots were obtained despite the fact that macroscopic ethanol droplets spread on the gold-coated glass slides. Possible explanations for this include the Laplace pressure at the concave menisci the ink forms at the mesopore openings on the stamp surface, and the reduced wettability of the 1-dodecanethiol dots themselves by ethanol that may impede ink spreading. The chemical heterogeneity of the surfaces thus patterned templated the dewetting of PS melts, eventually yielding ordered arrays of discrete PS nanoparticles. Beyond the application examples presented here, capillary nanostamping has the potential to enable deposition of drop arrays consisting of a broad range of inks on a likewise diverse range of substrates. Potential applications of capillary nanostamping may include the production of sensor arrays for bioanalytics as well as preconcentration sensing and of topographically or chemically patterned surfaces having tailored wettability or adhesive properties.

## 4. Experimental
### 4.1 Materials
Macroporous silicon (mSi; Supporting Figure S1) was obtained from SmartMembranes (Halle/Saale, Germany). Dimethyldichlorosilane (DMDCS) and octanol were obtained from Merck. *N*-hexane, chloroform, urea, Pluronic-F127 ($EO_{108}PO_{70}EO_{108}$), acetic acid, methyltrimethoxysilane (MTMS), potassium hydroxide and PS ($M_w$ = 32000 g/mol, $M_n$ = 31000 g/mol, $M_w/M_n$ = 1.02) were obtained from Sigma-Aldrich. Sylgard 184 formulation was obtained from Dow Corning. 1H,1H,2H,2H-perfluorodecyltrichlorosilane (FDTS), 1-dodecanethiol, sodium thiosulfate pentahydrate, potassium ferrocyanide (III) and potassium ferrocyanide (II) were obtained from Alfa Aesar. All chemicals mentioned above were used as received.
Fullerenes ($C_{60}$ : $C_{70}$ : higher fullerenes ≈ 78 wt-% : 20 wt-% : 2 wt-%) were obtained from New Jersey Institute of Technology (Newark, USA). A $C_{60}$ content exceeding 99.9 wt-% was achieved by sublimation at ~500 °C, chromatographic separation on activated charcoal/silica gel and preparative HPLC (Cosmosil Buckyprep column). Toluene (technical grade) was received from Sigma-Aldrich and purified by two successive distillations using a rotary evaporator.

### 4.2 Synthesis of silica-stamps using MTMS as silica precursor
Pieces of mSi with areas of 2 x 2 $cm^2$ were silanized by immersion into a mixture of 500 μL DMDCS and 9.5 g n-hexane at 40 °C for 24 h. The silanized mSi was then rinsed with ethanol and dried in air. For the synthesis of the mesoporous silica stamps, 14 g of 10 mM aqueous acetic acid solution were mixed with 2.2 g Pluronic F127 and 1 g urea in an Erlenmeyer flask and then vigorously stirred for 30 min at room temperature. Then, 9.97 mL MTMS were added quickly and stirring was continued for 30 min. Silanized mSi located in a polystyrene container was coated with the sol thus obtained. The container was tightly sealed, and the sol was allowed to age for 4 days at 60 °C. The obtained silica monoliths were neutralized with deionized water for 1 d at 60 °C to remove residual surfactant and chemicals. The water was replaced by ethanol, which was in turn replaced by fresh ethanol for two more times within 24 h. Finally, the mesoporous silica stamps were dried by critical point drying, which was carried out in a Leica Auto CPD300 with liquid $CO_2$ at 40 °C and 80 bar. Nitrogen sorption measurements were performed with a Porotec Surfer device at 77 K. Before any measurement, the samples were outgassed at 250 °C for 10 h.

### 4.3 Capillary nanostamping
An elastomeric PDMS film with a thickness of 3 mm was prepared as follows using Sylgard 184 formulation. Base and PDMS prepolymer were mixed at a ratio of 1:9, stirred for 10 min, and poured into a polyethylene mold. The mixture was then allowed to cure for 1 week at room temperature. As stamp holder we used a stainless steel cylinder with a height of 4.5 cm, a diameter of 2 cm, a mass of ~27 g and a flat cylinder base. Using double-sided adhesive tape, the elastomeric PDMS film glued onto the flat cylinder base of the stamp holder. The mesoporous silica stamps were loaded with ink as described below. Excess ink on the surfaces of the mesoporous silica stamps was removed with tissue. Then, the mesoporous silica stamps were fixated on the PDMS films with double-sided adhesive tape. The mesoporous silica stamps mounted on the stamp holders were manually brought into contact with glass slides modified as described below.
Fullerene nanoparticles were stamped on 0.17 mm thick glass slides extending 18 x 18 $mm^2$ supplied by VWR. Prior to capillary nanostamping, the glass slides were subjected to $O_2$ plasma for 10 min at an $O_2$ pressure of 5 mbar using a Diener femto plasma cleaner. The glass slides thus treated were then silanized with 1H,1H,2H,2H-perfluorodecyltrichlorosilane (FDTS) via vapor phase deposition. For this purpose, the glass slides were heated to 80°C for 2 h in a sealed glass container in the presence of an excess of FDTS (~2 µL). The mesoporous silica stamps were immersed into a solution of 1 mg $C_{60}$ per 1 mL toluene for ~5 minutes and then brought into contact with the FDTS-modified glass slides for ~5 s.

1-Dodecanethiol was stamped onto 1 mm thick glass slides extending 1 x 1 $cm^2$ supplied by VWR. Prior to capillary nanostamping, the glass slides were coated with titanium and gold by thermal evaporation in a vacuum chamber at $10^{-4}$ mbar using a Balzers BAE 120 evaporator. At first, we deposited a 5 nm thick titanium layer followed by the deposition of a 30 nm thick gold layer. To estimate the thickness of the titanium and gold layers, an AT-cut eQCM quartz having a defined area was coated with metal under the same conditions than the glass slides and weighed prior to and after metal deposition using a GAMRY eQCM 10M quartz crystal microbalance. Before capillary nanostamping, the metal-coated glass slides were rinsed with ethanol and dried in an argon flow. The mesoporous silica stamps were infiltrated by immersion into an ethanolic 2 mM 1-dodecanethiol solution for 5 minutes and then brought into contact with the gold-coated glass slides for ~1 s.

### 4.4 Dewetting of polystyrene

10 µL of solutions of PS in chloroform (concentrations see caption of Figure 7) were spin coated onto gold-coated glass slides modified with arrays of 1-dodecanethiol dots at 3000 rpm for 30 s using a spin coater G3P-8 from Specialty Coating Systems. To evaluate the thickness of the PS films, we spin-coated the solutions under exactly the same conditions onto glass slides without gold coating and measured the step height of scratches in the PS films with an atomic force microscope NT-MDT Ntegra. The PS films covering the gold-coated glass slides patterned with arrays of 1-dodecanethiol dots were first kept at 80°C, a temperature at which PS is vitreous, for 24 h to remove residual solvent. Then, the samples were heated to 170°C, a temperature well above the glass transition temperature of PS, for 30 min to enable dewetting. Note that we omitted the partial etching of the gold layer carried out to enhance the contrast for SEM imaging (see below); the gold layer on the glass slide was still intact during deposition and dewetting of the PS.

### 4.5 Characterization

Contact angles were measured in the sessile drop mode at 22°C and a humidity of 37 % using a drop shape analyzer DSA100 (Krüss, Germany). SEM was carried out on a Zeiss Auriga device at an accelerating voltage of 3 keV using a secondary electron chamber detector (SESI detector) and an in-lens detector. Prior to SEM investigations of gold-coated glass slides modified with arrays of 1-dodecanthiol dots, the gold layer in between of the 1-dodecanethiol dots was partially etched to enhance the SEM contrast by applying procedures reported elsewhere.[39]. For this purpose, we immersed the samples into a solution containing sodium thiosulfate, potassium ferrocyanide (III), potassium ferrocyanide (II), potassium hydroxide and octanol at a molar ratio of 1:0.1:0.01:10:0.02 for 6 min at room temperature. Subsequently, the substrates were rinsed with water as well as with ethanol and dried in an argon flow. The initial thickness of the gold layer amounted to 30 nm. After partial etching of the Au layer, atomic force microscopy measurements (Supporting Figure S14) revealed a peak-to-valley height of ~20 nm between the 1-dodecanethiol dots and their surroundings. Hence, the thickness of the remaining gold layer was estimated to be ~10 nm. Analysis of the SEM images was carried out using the program ImageJ. Brightness and contrast were adjusted in such a way that shape and size of the $C_{60}$ nanoparticles and the 1-dodecanethiol dots displayed in the SEM micrographs were invariant against small brightness and contrast changes. The SEM images were then converted into binary images.

**Supporting Information**
Supporting Information is available from the Wiley Online Library or from the author.

**Acknowledgements**
The authors thank the European Research Council (ERC-CoG-2014, project 646742 INCANA) for funding. Technical support by Jonas Klein (QCM measurements) and Xihomara Lizzet Casallas Cruz (contact angle measurements) is gratefully acknowledged.

**Conflict of Interest**
The authors declare no conflict of interest.

**Keywords**
Sol-gel chemistry, stamping, fullerenes, alkyl thiols, dewetting